\documentclass[12pt]{article}
\usepackage{amssymb}
\usepackage{amsmath}
\usepackage{amsfonts}
\usepackage[applemac]{inputenc}
\usepackage{seceqn}
\title{On the Wheeler-DeWitt equation for Kasner-like cosmologies}
 
\author {M. Maceda \\[8pt]
               Laboratoire de Physique Théorique\\
               Université de Paris-Sud, Bâtiment 211, F-91405 Orsay}

\date{}

\def\be{\begin{equation}}
\def\ee{\end{equation}}
\def\benum{\begin{enumerate}}
\def\eenum{\end{enumerate}}
\def\stk#1 #2{{\stackrel #1 #2}}

\def\t#1{\tilde #1}
\def\Dirac{{\raise0.09em\hbox{/}}\kern-0.69em D}

\def\tfrac #1#2{\textstyle{\frac{#1}{#2}}}

\begin{document} 

\maketitle

\begin{abstract}
 The Wheeler-DeWitt equation is obtained for Kasner-like cosmologies. Some solutions to this equation are presented for empty space, space filled with a cosmological constant and in the presence of a scalar field. We also briefly discuss a non-commutative extension of these results. 
\end{abstract}

\vfill

\newpage

\section{Introduction}

  The idea of quantizing gravity has been for long time being pursued and several approaches for this task have been developed. The canonical theory was one of the earlier attempts to follow this program and is based on a hamiltonian formulation of the field equations of general relativity~\cite{Wheeler:1964,Wheeler:1967,deWitt:1967}. One of the advantages of using a hamiltonian formulation is the fact that the notion of an absolute time does not exist, the time scale being arbitrary~\cite{Dirac:1964}, however the equations of motion do not suffer of ambiguities and they remain invariant under re-scaling.  
  
  However several difficulties were present in this approach: it was not possible to associate conjugate momenta to some of the variables used in this formulation and therefore primary and secondary constraints had to be considered. The final outcome of this work were the dynamical constraints $p_N \Psi = 0,\, H \Psi = 0$ on a state vector $\Psi$. The usual interpretation is that all the important information of the gravitational field that is coordinate-independent will be encoded on the hamiltonian constraint $H \Psi = 0$.   
 
 As in quantum mechanics, the wave function $\Psi$ is intended to determine completely the properties of physical systems of interest as quantum states evolving accordingly to the hamiltonian $H$. 
 Due to the technical difficulty in treating the problem in all its generality, one is then led to analyze cosmologies that are distinguished because of the description of the universe that they might provide. In this context the Friedmann-Robertson-Walker universe has been studied in the presence of scalar and matter fields ~\cite{Khvedelidze:2001tr}. 

On the other hand it has been argued that near a cosmological time singularity the behaviour of a space-time is that of an ordinary manifold in empty space~\cite{Lifshitz:1963ps}. The study of homogeneous space-times with a group $SU(2)$ of transformations acting on it has also provided some arguments in favor of this idea. One can also show that if any other fields are present they can be considered as small perturbations having as a background a vacuum solution. Because of the interest in knowing the evolution of the universe from its beginning as a singularity, the study of a Kasner-like cosmology using the quantization scheme provided by the canonical theory is then worthwhile.

This paper is organized as follows: in section 2 we obtain the lagrangian and hamiltonian description for a Kasner-like cosmology. We use these results in section 3 to obtain the corresponding Wheeler-DeWitt equation and some solutions are presented in the case of empty space. Section 4 deals with the solution to the constraints $p_N \Psi = 0, H \Psi = 0$ in the presence of a cosmological constant. The case of a scalar field is then discussed in section 5. We end with some general remarks on the extension to the noncommutative setting of these results.

\section{Classical description}

  We shall consider the following 4-dimensional Kasner-like metric 
$$
ds^2 = N^2 dt^2 - \sum_{i=1}^3 e^{2\alpha_i} (dx^i)^2
$$
where $N$ represents the lapse function and the $\alpha_i$'s are functions depending only on $t$. Using the frame components
$$
\theta^0 = N dt, \qquad \theta^i = e^{\alpha_i} dx^i
$$
the Cartan formalism gives after a straightforward calculation the following expression for the curvature scalar 
$$
R = - 2 N^{-2} \left (\sum_i (\ddot \alpha_i -\alpha_i N^{-1} \dot N + {\dot \alpha_i}^2) + \sum_{i < j} \dot \alpha_i \dot \alpha_j \right ).
$$
As in the usual way, the corresponding density Lagrangian from which we shall obtain the Wheeler-DeWitt equation is given by
\be
L = - \frac 1{16\pi G} R \sqrt{- g} = \frac 1N e^{\sum_i \alpha_i} \left ( \sum_i (\ddot \alpha_i -\alpha_i N^{-1} \dot N + {\dot \alpha_i}^2) + \sum_{i < j} \dot \alpha_i \dot \alpha_j \right )
\ee
where we have set $8\pi G = 1$.
  As discussed in~\cite{Reuter:1992}, if one imposes $N=1$ and afterwards one finds a solution to the variational equations $\delta L/ \delta \dot \alpha_i = 0$, it is not sure that the action will be stationary at it. 
 The appropriate way to proceed is to consider $N$ as a free parameter and set $N=1$ only after writing the dynamical constraints $p_N \Psi =0$ and $H \Psi = 0$. 

The second derivatives appearing in the previous expression can be eliminated by adding the total divergence
$$
L_{div} = - \frac d{dt} \left[ \frac 1N e^{\sum_i \alpha_i} \sum_i \dot \alpha_i  \right]  
$$
leading to the simple result
\be
L_T = - \frac 1N e^\zeta \sum_{i < j} \dot \alpha_i \dot \alpha_j, \qquad \zeta = \sum_i \alpha_i .	\label{lag}
\ee
The associated momenta
$$
\pi_{\alpha_i} = \frac{\partial L_T}{\partial \dot \alpha_i} 
$$
can readily be calculated, they are given by
$$
\pi_{\alpha_i} = - \frac 1N e^\zeta (\dot \alpha_2 + \dot \alpha_3, \dot \alpha_1 + \dot \alpha_3, \dot \alpha_1 + \dot \alpha_2).
$$
Using these results the Hamiltonian density can be written as
\be
H = \sum_i \dot \alpha_i \pi_{\alpha_i} - L_T = - \frac N4 e^{-\zeta} \left[ 2 \sum_{i<j} \pi_i \pi_j - \sum_i \pi_i^2  \right].	\label{ham}
\ee
Let us consider the constraint $p_N = \partial L_T / \partial N = 0$, which reads
\be
\sum_{i < j} \dot\alpha_i \dot\alpha_j = 0.	\label{ham-cond}
\ee
As one can easily verify, the Kasner solution
\be
\alpha_i = \frac 1{1 + a + a^2} (1 + a, a(1 + a), -a) \ln t + c_i \equiv p_i \ln t	 + c_i\label{kasner}
\ee
satisfies this constraint and thus the action we have just considered will be extremal at these values of the variables $\alpha_i$. 

The inclusion of a cosmological term is done without difficulty, following~\cite{Reuter:1992} one obtains
$$
H =  N \left(\Lambda e^{2\zeta} - \frac 14 \left[ 2 \sum_{i<j} \pi_i \pi_j - \sum_i \pi_i^2  \right] \right) e^{-\zeta}
$$
and after integration over a space-like hypersurface $t = const.$, one finds
\be
\tilde H =  N V \left(\Lambda e^{2\zeta} - \frac 14 \left[ 2 \sum_{i<j} \pi_i \pi_j - \sum_i \pi_i^2  \right] \right) \label{hamilt}
\ee
where $V$  is the associated 3-dimensional volume. Since this volume factor does not change at all the characteristics of the model, we shall disregard it in the next sections. We also have the following generalization of~(\ref{ham-cond})
\be
\sum_{i < j} \dot\alpha_i \dot\alpha_j = \Lambda.	\label{ham-condlamb}
\ee
This condition can be obtained by considering again the constraint $p_N = \partial L / \partial N = 0$, where now $L$ is the lagrangian formed from~(\ref{lag}) and $L_\Lambda = - \Lambda N e^\zeta$. 
We shall consider some solutions to this equation in the following section.

\section{Quantization: Wheeler-DeWitt equation}

The standard procedure when quantizing a system described by a hamiltonian $H$ consists in replacing the momenta $\pi$ by differential operators $-i\hbar \nabla$. This is a well known procedure from quantum mechanics, however, before doing that some words are in order. One of the ambiguities, if not the most important when translating quantities into their quantum counterparts is the fact that they do not longer commute. The Weyl prescription can be useful in this context, another possibility is to look for a kind of `natural' ordering associated to the problem which is under study but a satisfactory answer to this does not exist.     

We shall chose a simple factor ordering and substitute each of the momenta appearing in~(\ref{hamilt}) concordantly. This leads to the expression
\be
\hat H =  \frac {N\hbar^2}4 \left[  2 \sum_{i<j} \nabla_i \nabla_j - \sum_i \nabla_i^2  \right] + N \Lambda e^{2\zeta}, \qquad \nabla_i = \frac {\partial }{\partial \alpha_i}. 	\label{hamWdW}
\ee
%As mentioned before, we also have to take into account the constraint~(\ref{ham-condlamb}). 

The simplest case one can consider for the above equation is that of empty space, i.e. $\Lambda = 0$.
A solution to $\hat H \Psi = 0$ is then $\Psi $ being linear on the $\alpha_i$'s. This is a rather trivial solution that is also unbounded on the space defined by these variables. It is then interesting to look for other solutions that can be regular.

 Consider the Ansatz
$$
\Psi = e^{i\rho \alpha_1 + i \sigma \alpha_2 + B(\alpha_3)}.
$$
As a straightforward calculation shows, from the equation
\be
\left[  2 \sum_{i<j} \nabla_i \nabla_j - \sum_i \nabla_i^2  \right] \Psi = 0
\label{WdWempsp}
\ee
it follows that the function $B(\alpha_3)$ should satisfy the differential equation
\be
\ddot B + \dot B^2 - 2i(\rho + \sigma) \dot B - (\rho - \sigma)^2 = 0.
\label{diffeqb}
\ee
Therefore, after a suitable translation on $\dot B$ one has 
$$
\dot B = i(\rho + \sigma) - \mu \frac {c_0 e^{\mu\alpha_3} - c_1 e^{-\mu\alpha_3}}{c_0 e^{\mu\alpha_3} + c_1 e^{-\mu\alpha_3}}, \qquad \mu^2 = - 4\epsilon \rho\sigma > 0.
$$
In order to further proceed we shall set $c_0 = c e^\phi, c_1 = c e^{-\phi}$ with $\phi$ being interpreted as a constant phase. With this choice the fraction in the above expression becomes a hyperbolic function and a further integration gives then the final solution to~(\ref{diffeqb}) as
\be
B(\alpha_3) = i(\rho + \sigma) \alpha_3 - \ln |\cosh (\mu \alpha_3 + \phi) | + B_0.
\label{solb}
\ee
Inserting this solution in the expression for $\Psi$ one has
\be
\Psi = \Psi_0 e^{i\rho\alpha_1 + i\sigma\alpha_2  + i(\rho + \sigma) \alpha_3} / \cosh (\mu \alpha_3 + \phi).
\label{psisol}
\ee
Therefore the probability density is given by
\be
|\Psi|^2 = |\Psi_0|^2 / (\cosh (\mu \alpha_3 + \phi))^2. 	\label{prob}
\ee
Some words are in order: first, were $\mu$ negative then in~(\ref{solb}) instead of $\cosh$ one would have $\cos$ and therefore the density probability $|\Psi|^2$ would be divergent at the points $\t \alpha_3$ satisfying $\mu \t \alpha_3 + \phi = (2n+1) \pi/2$, with $n$ an integer. The fact that~(\ref{prob}) is bounded and vanishing when $\alpha_i \to \pm \infty$ points out to the possibility that regularized solutions can exist for more general situations other than empty space.
Another important point concerns the dependence on $\alpha_3$. As can be seen from~(\ref{hamWdW}), the Hamiltonian is invariant under a cyclic permutation of the indexes $1,2$ and $3$. It follows then that expressions similar to~(\ref{prob}) with $\alpha_1$ or $\alpha_2$ instead of $\alpha_3$ exist. 
%Unfortunately there is not solution of this kind involving the three $\alpha_i$'s due to the non-linearity of the equations. 

It is a straightforward calculation to find the current $\vec J = \frac {i\hbar}2 (\Psi \partial \Psi^* - \Psi^* \partial \Psi )$ associated to~(\ref{psisol}). One has
$$
\vec J = \hbar |\Psi|^2 (\rho, \sigma, \rho + \sigma )
$$
which is also bounded.

\section{Cosmological constant}\label{anscosm}

Before solving the Wheeler-DeWitt equation~(\ref{hamWdW}) in more generality, we shall consider the equation of motion for the variables $\alpha_i$, namely~(\ref{ham-condlamb}). In order to solve it we take the following Ansatz  
$$
\alpha_i (t) = p_i \ln \frac {F(t)}{F(t_0)} + q_i K(t)
$$
with the $p_i$'s and $q_i$'s being for the moment arbitrary constants.
Here $t_0$ is an arbitrary time to be specified later on and that plays the role of a scaling factor. Using this expression one arrives to the equation
\be
p^2 \frac {\dot F^2}{ F^2} + m^2 \frac {\dot F \dot K}F + q^2 \dot K^2 = \Lambda 
\label{genkasn}
\ee
where the quantities
$$
 p^2 \equiv \sum_{i<j} p_i p_j, \qquad m^2 \equiv \sum_{i<j} (p_i q_j + p_j q_i),  \qquad q^2 \equiv \sum_{i<j} q_i q_j
$$
have been defined.

We know that in the case of a vanishing cosmological constant the constants $p_i$ can be chosen to be the Kasner exponents once one sets $F(t) = \ln t$ and $K(t) = 1$ (see~(\ref{kasner})). In the following we would like to consider the departures from this ``classical" solution for small values of the cosmological constant. For this we shall solve $K(t)$ in terms of $F(t)$ using~(\ref{genkasn}).

There are two different cases to be considered depending on the value of $q^2$:

{\it a)} $q^2 = 0$

There is not quadratic term on $\dot K$ in~(\ref{genkasn}) and one can write directly
\be
K(t) = K(t_0) - \frac {p^2}{m^2} \ln \frac {F(t)}{F(t_0)} + \frac {\Lambda}{m^2} \int_{t_0}^t \frac F{\dot F} dt.
\ee
The solution is unique and one then has
$$
e^{\alpha_i} = (F(t)/F(t_0))^{p_i - q_i \frac {p^2}{m^2} } \exp \left[ K(t_0) + \frac {q_i \Lambda}{m^2} \int_{t_0}^t \frac F{\dot F} dt \right].
$$

{\it b)} $q^2 \neq 0$

In this case there are two different solutions 
\be
K_\pm (t) = K_\pm (t_0) -\frac {m^2}{2q^2} \ln \frac {F(t)}{F(t_0)}  \pm \int_{t_0}^t \left[ \frac \Lambda{q^2} + \frac {m^4- 4p^2 q^2}{4q^4} \frac {\dot F^2}{ F^2} \right]^{1/2} dt.
\ee
Accordingly we have
$$
e^{\alpha_{i\pm}} = (F(t)/F(t_0))^{p_i - q_i \frac {m^2}{2q^2} } \exp \left[ K_\pm (t_0) \pm q_i \int_{t_0}^t \left[ \frac \Lambda{q^2} + \frac {m^4- 4p^2 q^2}{4q^4} \frac {\dot F^2}{ F^2} \right]^{1/2} dt \right]
$$

Let us now specialize these solutions to the Kasner case, namely $p^2 = 0$ and $F(t) = t$. We also choose $t_0 = 1$ so that $F(t_0) = 1$. Case $a)$ gives then 
$$
K(t) = K(1) + \frac 12 \frac \Lambda{m^2} (t^2 - 1).
$$
By setting $K(1) = \Lambda/ 2m^2$ one has for $\Lambda \ll 1$ 
$$
e^{\alpha_i} = t^{p_i} (1 + \frac 12 \frac {q_i \Lambda}{m^2} t^2 )
$$
up to linear terms. It follows that to lowest order the volume element is then given by
\be
dV_{(a)} = \sqrt{ -g } d^4 x = N t ( 1 + \frac 12 \frac {r \Lambda}{m^2} t^2 ), \qquad r \equiv \sum_i q_i.
\label{vola}
\ee
In the above expression we have used the fact that $\sum_i p_i =1$.

For case $b)$ one has
\begin{equation*}
\begin{split}
K_\pm (t) &= K_\pm (1) - \frac {m^2}{2q^2} \ln t \pm  \frac {m^2}{2q^2} \int_1^t \left[ 1+ \frac {4 q^2 \Lambda}{m^4} t^2 \right]^{1/2} \frac {dt}t \\[6pt] 
&= K_\pm (1) - \frac {m^2}{2q^2} \ln t \pm  \frac {m^2}{2q^2} \left[ (1 + s^2)^{1/2} - \ln \left| \frac {1 + (1 + s^2)^{1/2} }s \right| - 2^{1/2} + \ln(1+2^{1/2}) \right]
\end{split}
\end{equation*}
where $s^2 = 4q^2 \Lambda t^2 /m^4$. For small $\Lambda$ we have
$$
K_+ (t) = \frac 12 \frac \Lambda{m^2} t^2, \qquad K_- (t) = - \frac {m^2}{q^2} \ln t - \frac 12 \frac \Lambda{m^2} t^2
$$
where we have chosen 
$$
K_+ (1) = - K_- (1) = - \frac {m^2}{2q^2}[1 + \ln |q \Lambda^{1/2}/m| - 2^{1/2} + \ln (1+2^{1/2}) ] .
$$
Using the above results one obtains
$$
e^{\alpha_{i+}} = t^{p_i} (1 + \frac 12 \frac {q_i \Lambda}{m^2} t^2 ), \qquad  e^{\alpha_{i-}} = t^{p_i - q_i \frac {m^2}{q^2} } (1 - \frac 12 \frac {q_i \Lambda}{m^2} t^2 ),
$$
and therefore the corresponding volume elements are
\be
dV_{(b)+} = N t ( 1 + \frac 12 \frac {r \Lambda}{m^2} t^2 ), \qquad dV_{(b)-} = N t^{1 - \frac {r m^2}{q^2}} ( 1 - \frac 12 \frac {r \Lambda}{m^2} t^2 )
\label{volb}
\ee

We see from~(\ref{vola}) and~(\ref{volb}) that to lowest order on $\Lambda$, the volume elements $dV_{(a)}$ and $dV_{(b)+}$ are equal while $dV_{(b)-}$ has a supplementary dependence on $t$ (the values of $K(1), K_\pm(1)$ have been chosen as to eliminate all the irrelevant constant scaling factors from these expressions). We can now define the ratio $\eta_i = dV_{(i)} /dV_K$, with $dV_K = Nt$ the volume element associated to the Kasner cosmology, in order to compare the different geometries. We have
\be
\eta_a = \eta_{b+} = 1 + \frac 12 \frac {r \Lambda}{m^2} t^2, \qquad
\eta_{b-} = t^{- \frac {r m^2}{q^2}}(1 - \frac 12 \frac {r \Lambda}{m^2} t^2 ).
\ee
It is clear that different behaviors in the limit $t \to 0$ are present: in the first case one can effectively replace $dV_{(a)}$ or $dV_{(b)+}$ by $dV_K$. The second case is more interesting since due to the presence of the exponent $z = rm^2/q^2$ the ratio $\eta_{(b)-}$ can either vanish ($z < 0$) or diverge ($z > 0$).

We shall now turn our attention to~(\ref{hamWdW}). After a translation $\alpha_i \to \alpha_i + c_i$ with $\hbar^2/4 = \Lambda e^{2 \sum c_i}/3$, it becomes
\be
\left[ 2 \sum_{i<j} \nabla_i \nabla_j - \sum_i \nabla_i^2 + 3 e^{2\zeta} \right] \Psi (\alpha_i) = 0.
\label{WdWred}
\ee
A solution can be obtained by considering $\Psi = \Psi (\zeta)$. Since $\nabla_i = \nabla_\zeta$ one arrives to the differential equation
$$
( \frac {d^2}{d{\zeta}^2} + e^{2\zeta} ) \Psi = 0.
$$
The equation can be solved by setting $x = e^\zeta$. Using this variable one obtains a Bessel differential equation of order 0. The solution is then simply given by 
\be
\Psi = A J_0 (x) + B Y_0 (x)
\ee
with $J_0(x), Y_0(x)$ the Bessel and Neumann functions of order 0 respectively. We have the following asymptotic developments
$$
\Psi(x) \sim \frac 1{\sqrt x} ( A \cos( x - \pi/4) + B \sin( x - \pi/4 ) ), \qquad x \to \infty
$$
and 
$$
\Psi(x) \sim A + B \ln (x/2), \qquad x \to 0.
$$
As in the case of empty space, a bounded solution for all values of $\zeta$ ($x \in [0, \infty)$) is possible to exist:  it suffices to take $A \neq0, B=0$.

\section{Inclusion of a scalar field}\label{gab}

When massless scalar and vector fields are present, it is known that the behavior of the Kasner exponents in the classical solution is modified~\cite{Belinskii:1973}. Through evolution in time, the classical oscillatory regime, known as Kasner epochs~\cite{Belinskii:1972} and characterized by having one of the $p_i$'s to be negative is modified and in fact, a final Kasner epoch is reached where all the $p_i$'s take positive values. The reason for this is that in the early stages of the cosmic evolution, a massless scalar field $\phi$ will give a contribution not negligible to the total stress-energy tensor, roughly of the same order as the gravitational interaction.

The kinetic energy associated to a scalar field $\Phi$ depending only on $t$ is given by
$$
\frac 12 \int \sqrt{-g}  g^{\mu\nu} \Phi_{,\mu} \Phi_{,\nu} d^4 x =  \frac V{2N} \int e^\zeta \dot \Phi^2 dt
$$
and therefore its contribution to the quantized hamiltonian~(\ref{hamWdW}) reads
$$
\hat H_\Phi = \frac N2 \Pi_\Phi^2 = - \frac {\hbar^2 N}2 \nabla_\Phi^2.
$$
The full Wheeler-DeWitt equation to be considered is then
\be
 \left \{ \frac {\hbar^2}2 \left[   \sum_{i<j} \nabla_i \nabla_j - \frac 12 \sum_i \nabla_i^2 - \nabla_\Phi^2 \right] + V(\Phi) e^{2\zeta} \right\} \Psi = 0.
\ee
We rewrite the first two terms in the brackets with the help of a metric $g^{AB}$ as
$$
\sum_{i<j} \nabla_i \nabla_j - \frac 12 \sum_i \nabla_i^2 = g^{AB} \nabla_A \nabla_B, \qquad
g^{AB} = \frac 12
       \begin{pmatrix}
       -1 & 1  & 1\\
       1  & -1 & 1\\
       1  & 1  & -1 \end{pmatrix}.
$$
Through the usual trick of diagonalization one has  
\be
\left \{ \frac 12 \left[  - \nabla_{x_1}^2 - \nabla_{x_2}^2 + \frac 12 \nabla_{x_3}^2 - \nabla_\Phi^2 \right] + V(\Phi) e^{2\zeta} \right\} \Psi = 0
\label{WdWx}
\ee
where now
$$
\zeta = \frac 1{\sqrt 3} [ (1+x) x_1 + (1+x) x_2 + (1-2x) x_3], \qquad x = \sqrt{2/3}.
$$
We have also performed a similar translation as in~(\ref{WdWred}) in order to eliminate a factor $\hbar^2$ (for details see the Appendix). 

The above transformation is suitable to find some solutions to~(\ref{WdWx}) both for $V(\Phi) = 0$ and $V(\Phi) = \Lambda$. We shall not dwell on this, instead we shall concentrate on the solutions to the classical equations of motion in the presence of the scalar field $\Phi$. In order to do so we shall stick to the variables $\alpha_i$ since they are better adapted to make the connection with the Kasner metric.
We have accordingly
$$
\sum_{i<j} \dot \alpha_i \dot \alpha_j = \frac 12 \dot \Phi^2 + V(\Phi), \qquad 
\ddot \Phi + \dot \zeta \dot \Phi = - \frac {dV}{d\Phi}. 
$$
We remark that the problem contains now four unknown functions to be determined, however we only have at our disposal two equations. Instead of trying to find the most general solution we shall take a similar Ansatz as that of section~\ref{anscosm}
\be
\alpha_i (t) = p_i \ln t + q_i K(t)
\ee
with $p_i$ being a Kasner exponent. We shall also set $V(\Phi) = \Lambda$. By doing this one arrives to
$$
m^2 t \dot K + q^2 t^2 \dot K^2 = \frac {c^2}2 e^{-2 r K} + \Lambda t^2, \qquad \dot \Phi = c t^{-1} e^{-r K}
$$
Making the change of variable $s = e^{2 r K}$ one finally obtains
\be
\frac {m^2}{2r} t s \dot s + \frac {q^2}{4r^2} \dot s^2 = \frac {c^2}2 s + \Lambda t^2 s^2.
\label{diffeqs}
\ee
In this last equation we have also supposed $r \neq 0$ (otherwise with some little modifications we can use directly the results of the precedent section to write down a solution).

Unfortunately we have been not able to find the most general solution to~(\ref{diffeqs}). However, for the special case $q^2 = 0$ one obtains
\be
s(t) = c_1 e^{r\Lambda t^2/m^2} - \frac {rc^2}{2m^2} e^{r\Lambda t^2/m^2} \,\mbox{Ei} (1, r\Lambda t^2/m^2).
\ee
The function $\mbox{Ei} (n,x)$ can be written as an infinite series on $x$ and is ill-defined at $x=0$, it has a logarithmic divergence. Due to the fact that $\dot \Phi \sim t^{-1} s^{-1/2}$, it follows also that for this solution the scalar field $\Phi$ can not be defined for values of $t$ smaller than a cut-off $t_0 > 0$. 

\section{Conclusions}

  We have found the Wheeler-DeWitt equation for 4-dimensional Kasner-like space-times and some cases of interest have been considered. For empty space the main result is that well-defined solutions exist in the space of the parameters of the model and the equations of motion admit the well known Kasner metric as solution. When a potential corresponding to a cosmological constant is introduced, a simple quantum state $\Psi$ can be found which has finite density probability. The solution to the classical equations of motion for this case show that different behaviors for the geometry of the space are present in the limit $t \to 0$. With a proper choice of the parameters involved the volume element of each solution can be made bigger than that of the Kasner metric. In particular when $0<z<1$, the ratio $\eta_{b-}$ diverges. 

The presence of a scalar field presents some additional complications. However, to write down a explicit solution should not present any special difficulties using equation~(\ref{WdWx}). On the other hand, the classical equations of motion allow for a solution for the scalar field in a simplified case. However the solution thus found is defined only in a certain region ($t > t_0$) and %. This solution %, though far of being unique, 
shows the kind of problems one may encounter in the most general case.    

A generalization of the above results to dimensions greater than 4 should be possible in a straightforward way. We would like to end with some comments on a non-commutative extension of these results. The interest in introducing a non-commutative structure for `classical' theories in some respect comes from the natural generalization it provides. From the physical point of view this has been important in giving new insight to some old problems~\cite{Grosse:2004ik, Bellisard:1994}. Here we are mainly interested in the link with gravity that can be stablished~\cite{Madore:1997ec} and the natural cut-off, essentially of the order of Planck's length, that it might provide to regularize singularities either space or time-like~\cite{Madore:1997ta, Maceda:2003xr}. 
  
  Examples of the role non-commutativity can play in cosmology are to be found in~\cite{Garcia-Compean:2001wy,Beciu:2003ps, Barbosa:2004kp}. The most natural way to introduce non-commutativity is through the use of the Moyal product, which in its differential form is given by~\cite{Osborn:1994sf}
$$
(f \star g) (x) = \exp \left [ \frac i2 \theta^{\mu\nu} \partial_\mu \partial_\nu^\prime \right ] \, f(x) g(x^\prime) |_{x = x^\prime}.
$$
The non-commutative parameter $\theta$ is defined through the commutation relations $[\alpha_i, \alpha_j] = i\theta_{ij}$. With this one can then write 
$$
\left[  2 \sum_{i<j} \nabla_i \nabla_j - \sum_i \nabla_i^2  + V_T  e^{2\zeta}  \right] \star \Psi (\alpha_i) = 0
$$
where $V_T$ is an arbitrary potential. The non-commutativity of the $\alpha_i$'s can be translated into a redefinition of these variables by making the replacement $\alpha_i \to \alpha_i + \frac i2 \theta_{ij} \pi^j$. This change will affect only the term $V_T e^{2 \zeta}$. An account of the consequences of this will be presented elsewhere.

\appendix

\section*{Appendix}
\setcounter{section}{1}
From the metric $g^{AB}$ introduced in section~\ref{gab} we have
$$
g_{AB} = 
       \begin{pmatrix}
       0 & 1  & 1\\
       1  & 0 & 1\\
       1  & 1  & 0 \end{pmatrix}.
$$
The eigenvalues of $g_{AB}$ are $-1$ (multiplicity 2) and $+2$. The matrix diagonalizing $g_{AB}$ is
$$
D^{-1} = \frac 1{\sqrt 3}
       \begin{pmatrix}
       2x & -x  & -x\\
       -x  & 2x & -x\\
       1  & 1  & 1 \end{pmatrix}, \qquad
x = \sqrt {2/3}
$$
and hence $\t g_{AB} = (D^{-1} g D)_{AB} = \mbox{diag} (-1, -1, 2)$. It follows that
$$
\t \nabla_A = \t g_{AB} \t \nabla^B = \t g_{AB} (D^{-1})^B_N g^{NM} \nabla_M = (D^{-1})^B_A \nabla_B.
$$
Explicitly we have
\be
\begin{array}{l}
\t \nabla_{x_1} = \tfrac x{\sqrt 3} [ 2 \nabla_1 - \nabla_2 - \nabla_3 ] \\[6pt]
\t \nabla_{x_2} = \tfrac x{\sqrt 3} [ - \nabla_1 + 2 \nabla_2 - \nabla_3 ] \\[6pt]
\t \nabla_{x_3} = \tfrac 1{\sqrt 3} [  \nabla_1 + \nabla_2 + \nabla_3 ]
\end{array}
\ee
and
\be
x_1 = \tfrac 1{\sqrt 2} [ \alpha_1 + x \alpha_3 ] \qquad
x_2 = \tfrac 1{\sqrt 2} [ \alpha_1 + x \alpha_3 ] \qquad
x_3 = \tfrac 1{\sqrt 2} [ -\alpha_1 - \alpha_2 + x \alpha_3 ].
\ee

%\setlength{\parskip}{5pt}
%\bibliographystyle{/Users/maceda/Documents/These/utphys}
%\bibliography{//Users/maceda/Documents/These/refart}
  
\providecommand{\href}[2]{#2}\begingroup\raggedright\endgroup  

\end{document}